\documentclass[a4paper]{JAC2003}
\title{BEAM--BEAM AND IMPEDANCE\thanks{Work partially supported by Brookhaven Science Associates, LLC under Contract
No. DE-AC02-98CH10886 with the U.S. Department of Energy, and in part by the U.S. LHC Accelerator Research Program.}}

\author{S. White, Brookhaven National Laboratory, Upton, NY, USA\\
X. Buffat, EPFL, Lausanne, Switzerland; CERN, Geneva, Switzerland}

\usepackage{graphicx}
\usepackage{booktabs}
\usepackage{varwidth}
\usepackage{xcolor}

\begin{document}
\maketitle

\begin{abstract}
As two counter-rotating beams interact they can give rise to coherent dipole modes. Under the influence of impedance these coherent beam--beam modes can couple to higher order head--tail modes and lead to strong instabilities. A fully self-consistent approach including beam--beam and impedance was used to characterize this new coupled mode instability and study possible cures such as a transverse damper and high chromaticity.
\end{abstract}

\section{Introduction}

In a high-energy, high-brightness hadron collider, the coherent dynamics of colliding beams is dominated by beam--beam interactions. The non-linearities of the beam--beam force introduce a tune spread largely exceeding the one from other sources, such as non-linear fields, and provide sufficient stability for any pure impedance instabilities. When they interact with each other, the two beams will couple, resulting in coherent oscillations. In the case of equal beams and tunes these coherent oscillations can be described by two eigenmodes, corresponding to either in-phase or out-of-phase oscillation, respectively the $\sigma$- and $\pi$-modes. The frequency of these modes may be well separated from the incoherent tune spread and consequently they do not profit from the large intrinsic Landau damping properties of the beam--beam interactions. Such coherent beam--beam modes are generally not self-excited and require some external mechanism to become unstable, such as the machine impedance. When studying the stability of colliding beams, it is therefore necessary to consider both processes simultaneously.

Past studies have shown that the combination of beam--beam interactions and impedance could lead to coherent instabilities. However, these studies were performed either using a linearized model \cite{ht_bb_circ} or for very specific cases applied to the Tevatron \cite{Tev_stern}. During the 2012 proton run of the LHC, coherent instabilities were routinely observed \cite{obs_LHC}, triggering a renewed interest to pursue these studies. In this paper, we present a refined model allowing one to study the interplay of beam--beam and impedance using the full LHC impedance model \cite{mounet_PhD}. We will concentrate mostly on single-bunch effects and associated stabilization techniques and present preliminary results for multibunch effects.

\section{Models}

Two models were used to characterize the interplay of beam--beam and impedance:

\begin{itemize}
 \item An analytical model based on the circulant matrix approach used in Ref. \cite{ht_bb_circ}
 \item A fully self-consistent multiparticle tracking model. Single-bunch effects were studied with the code BeamBeam3D \cite{BB3D} and multibunch effects with the code COMBI \cite{pieloni_PhD}
\end{itemize}

The circulant matrix model (CMM) allows one to compute the complex tune shift in the presence of six-dimensional beam--beam interactions, impedance, chromaticity and transverse damper. The bunches are sliced in the longitudinal phase space and the beam--beam kicks are computed with the linearized approximation. 
Landau damping is not included in this model but the computation of the eigenmodes is very fast allowing for extensive parameter scans. This provides an excellent tool for understanding the coherent beam dynamics in the presence of various physics processes.
This approach is very fast and most appropriate for extensive parameter scans, which provide a good understanding of the coherent dynamics in the presence of various processes. \\
While much more demanding in terms of computing power, tracking simulations are a necessary complement to the CMM. Indeed, the CMM is not a self-consistent approach, giving rise to differences in the frequencies of beam--beam coherent modes. In simple cases, the beam--beam parameter can be re-scaled in order to compensate for the change of frequency of the modes due to lack of self-consistency. In more complex configurations, this approximation needs to be tested against a self-consistent model. Also, the CMM is not suited to study any non-linear effect, in particular Landau damping. Indeed, even though the complex tune shift can be computed, the dispersion integral used to derive the stability of a pure impedance mode is not valid for a beam--beam coherent mode. An analytical derivation of a dispersion integral in LHC-type configurations promises to be a great challenge, in particular in the multibunch regime, i.e. in the presence of PACMAN effects. A numerical approach, by the means of self-consistent multiparticle tracking codes, allows one to address these issues. Moreover, such approach allows one to treat any other non-linear effects, e.g. transverse feedback imperfections or external noise. BeamBeam3D and COMBI (COherent Multi-Bunch Instabilities) are two similar implementations of such model, based on different multicore parallelization concepts. \\

\begin{figure}[htb]
\begin{center}
\includegraphics[width=0.45\textwidth]{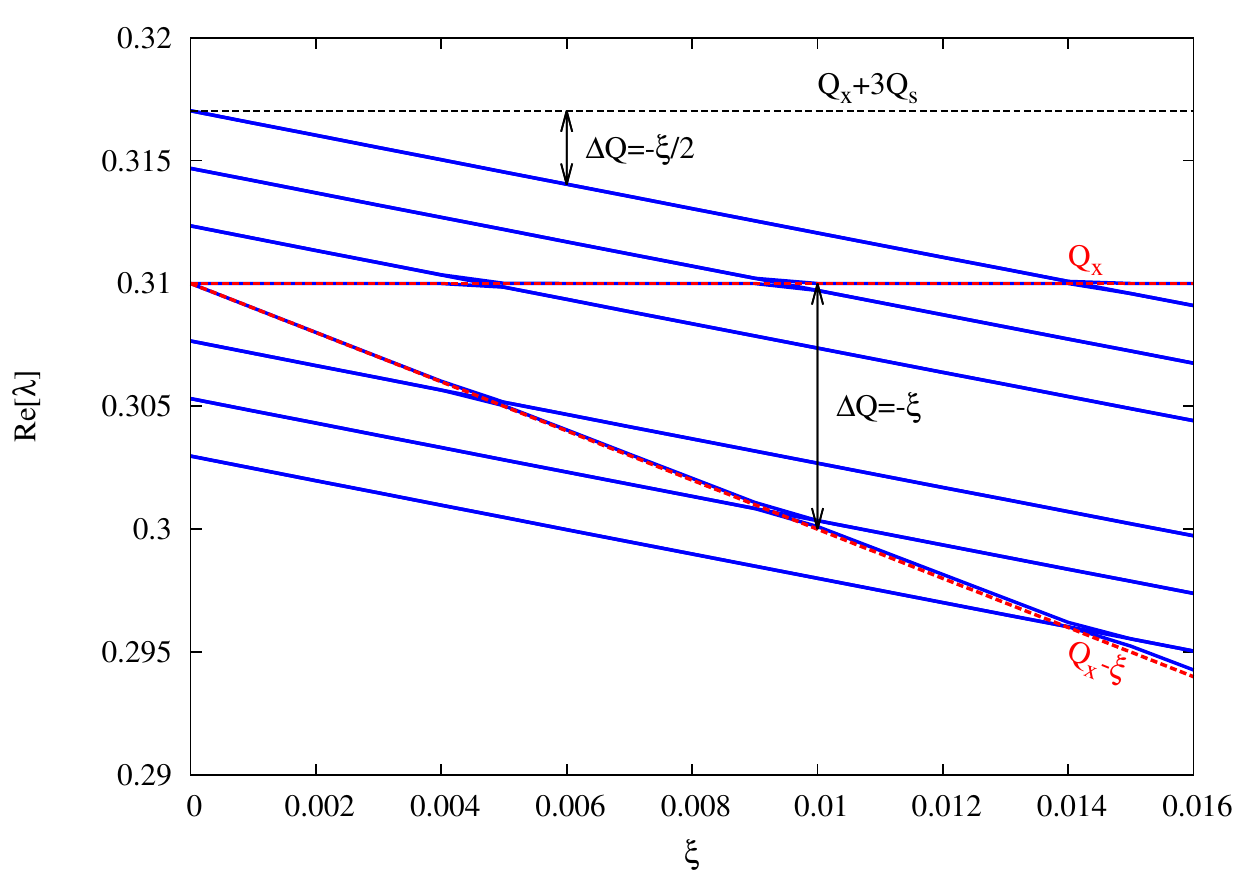}
\end{center}
\caption{Synchro--betatron modes as a function of the beam--beam parameter for $Q'=0.0$ and $\beta^*/\sigma_s\approx100$. Impedance was not included in this case.
The $\sigma$- and $\pi$-modes are shown in red.}
\label{bb6D_100}
\end{figure}

\begin{figure}[htb]
\begin{center}
\includegraphics[width=0.45\textwidth]{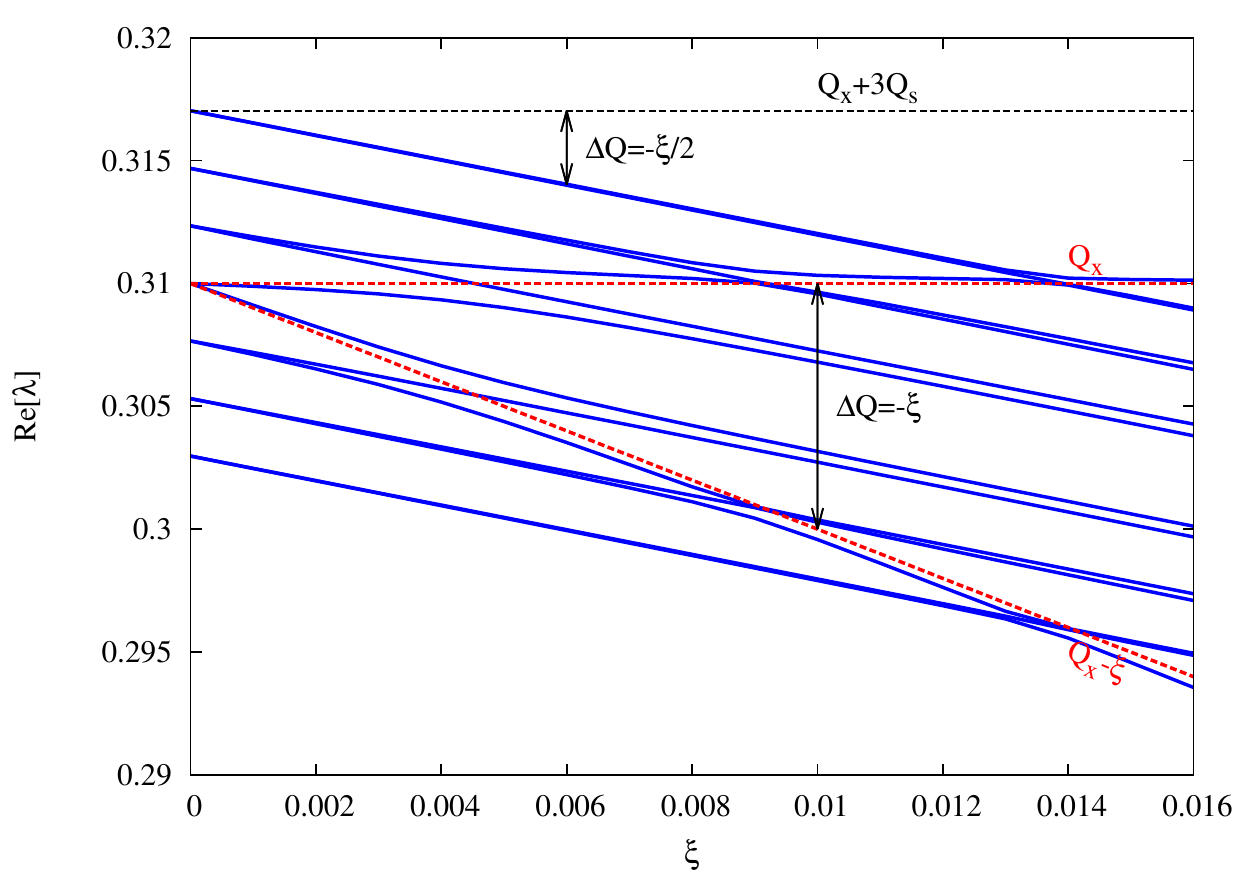}
\end{center}
\caption{Synchro--betatron modes as a function of the beam--beam parameter for $Q'=0.0$ and $\beta^*/\sigma_s\approx1$. Impedance was not included in this case.
The $\sigma$- and $\pi$-modes are shown in red.}
\label{bb6D_1}
\end{figure}

Figures \ref{bb6D_100} and \ref{bb6D_1} show the tunes of the synchro--betatron modes up to the third sideband in the presence of a six-dimensional beam--beam interaction. When the synchro--betatron coupling introduced by the beam--beam interaction is negligible ($\beta^*/\sigma_s>>1$), there is no cross talk between higher order head--tail modes and the coherent beam--beam $\sigma$- and $\pi$-modes. The tune of the sidebands is shifted by the coherent beam--beam tune shift, which is approximately equal to $\xi/2$ in this case. When the synchro--betatron coupling becomes important ($\beta^*/\sigma_s\approx1$), the synchrotron sidebands are now deflected when their frequency approaches the frequency of the coherent beam--beam modes, indicating possible coupling between the coherent beam--beam dipolar modes and higher order head--tail modes. In both cases, the imaginary part of the tune shifts of all modes is equal to zero: in the presence of beam--beam interactions only, the system is always stable.

\begin{figure}[htb]
\begin{center}
\includegraphics[width=0.45\textwidth]{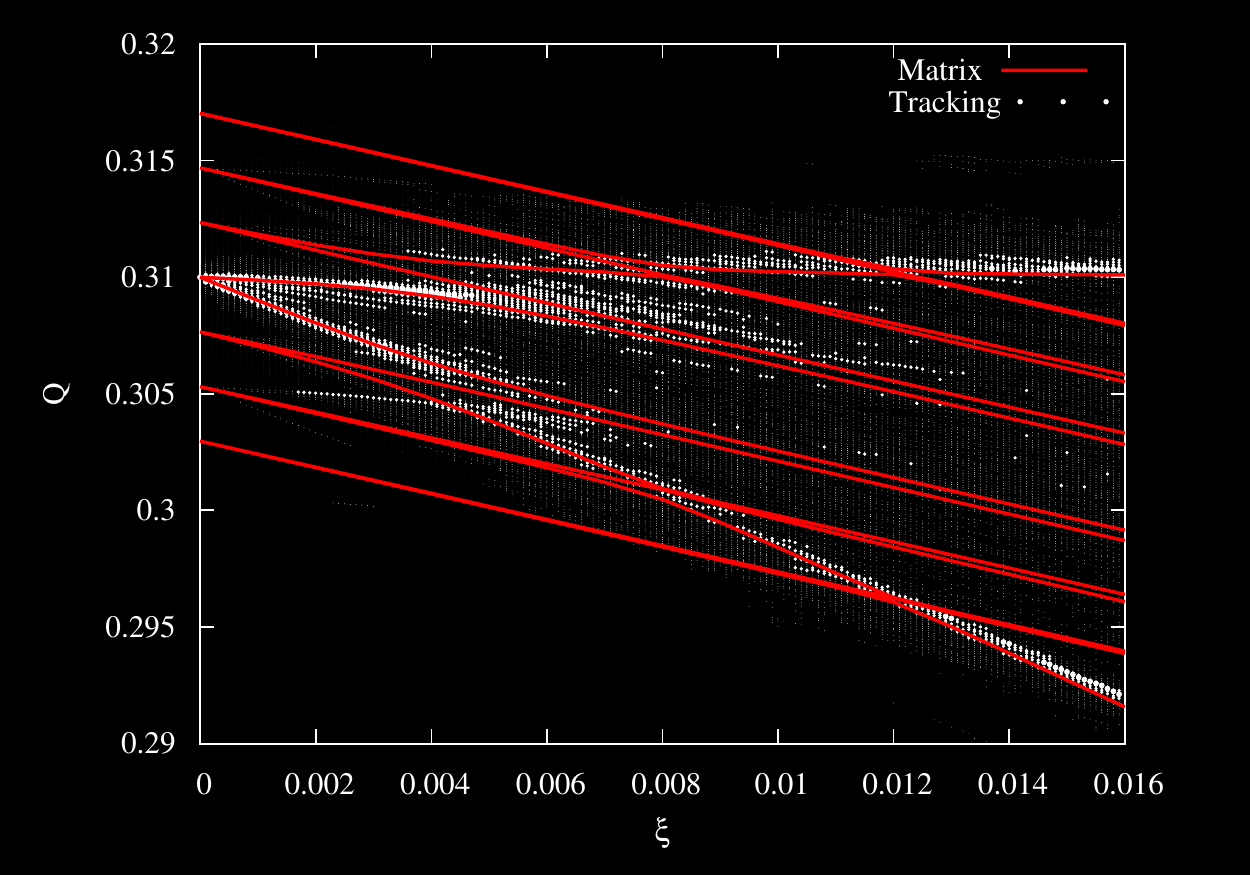}
\end{center}
\caption{Synchro--betatron modes as a function of the beam--beam parameter for $Q'=0.0$ and $\beta^*/\sigma_s\approx1$. Comparison between the CMM
and BeamBeam3D. Impedance was not included in this case.}
\label{bb6D_track}
\end{figure}

Figure \ref{bb6D_track} shows a comparison between the tracking code BeamBeam3D and the CMM for $\beta^*/\sigma_s\approx1$. The CMM was re-scaled by the Yokoya factor to match the tracking results. An excellent agreement is observed and one can see that the frequency of the modes is modified when the beam--beam coherent modes cross the first sidebands. The frequency components between the $\sigma$- and $\pi$-modes observed in the tracking correspond to the beam--beam tune spread.

\begin{figure}[htb]
\begin{center}
\includegraphics[width=0.45\textwidth]{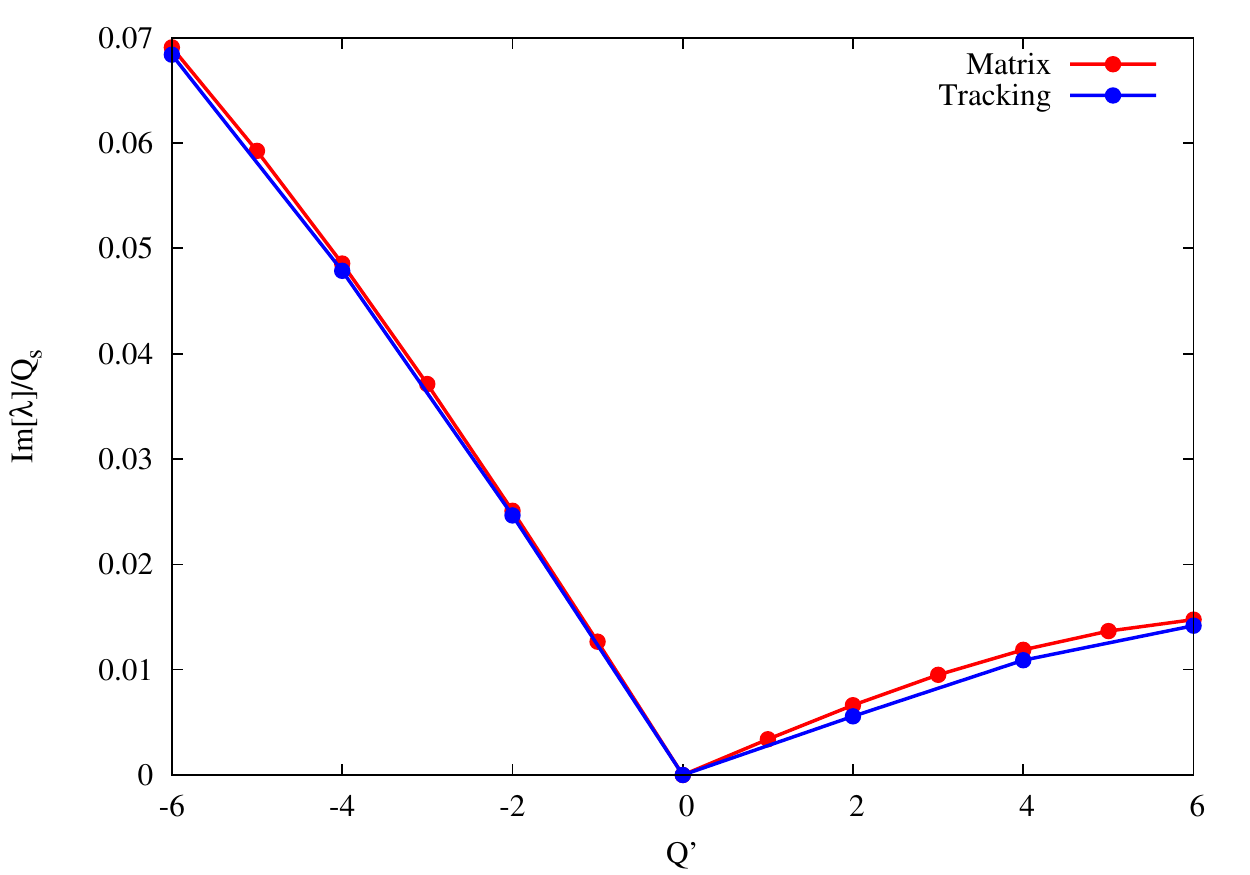}
\end{center}
\caption{Imaginary part of the tune shift as a function of chromaticity for the CMM and BeamBeam3D. In both models an airbag distribution
was used.}
\label{comp_rt}
\end{figure}

The implementation of impedance in BeamBeam3D was fully benchmarked with the HEADTAIL code \cite{headtail}. In order to validate the implementation of the LHC impedance model into the the CMM, we compared the rise times as a function of chromaticity for an airbag distribution. The results are shown in Fig. \ref{comp_rt}, where an excellent agreement is observed. The implementation of the LHC impedance model in COMBI was also benchmarked against multibunch HEADTAIL \cite{mounet_PhD}; nevertheless, this development is rather recent and only preliminary results are presented.

\section{Mode coupling instability of colliding beams}

We start by looking at the simple case of two single bunches colliding head-on in one interaction point (IP). The impedance model used in the following simulations was derived using the 2012 LHC lattice and collimator settings \cite{mounet_PhD}. The beam--beam interactions are computed with a full six-dimensional model taking into account the synchro--betatron effects and eventual non-Gaussian transverse distributions. In order to estimate the beam stability for a large number of beam parameters, multiparticle tracking is performed over 10000 turns and each case is analysed using an interpolated FFT algorithm. The beam stability of any given mode can then be assessed by looking at the amplitude of its corresponding spectral line.

\begin{figure}[htb]
\begin{center}
\includegraphics[width=0.45\textwidth]{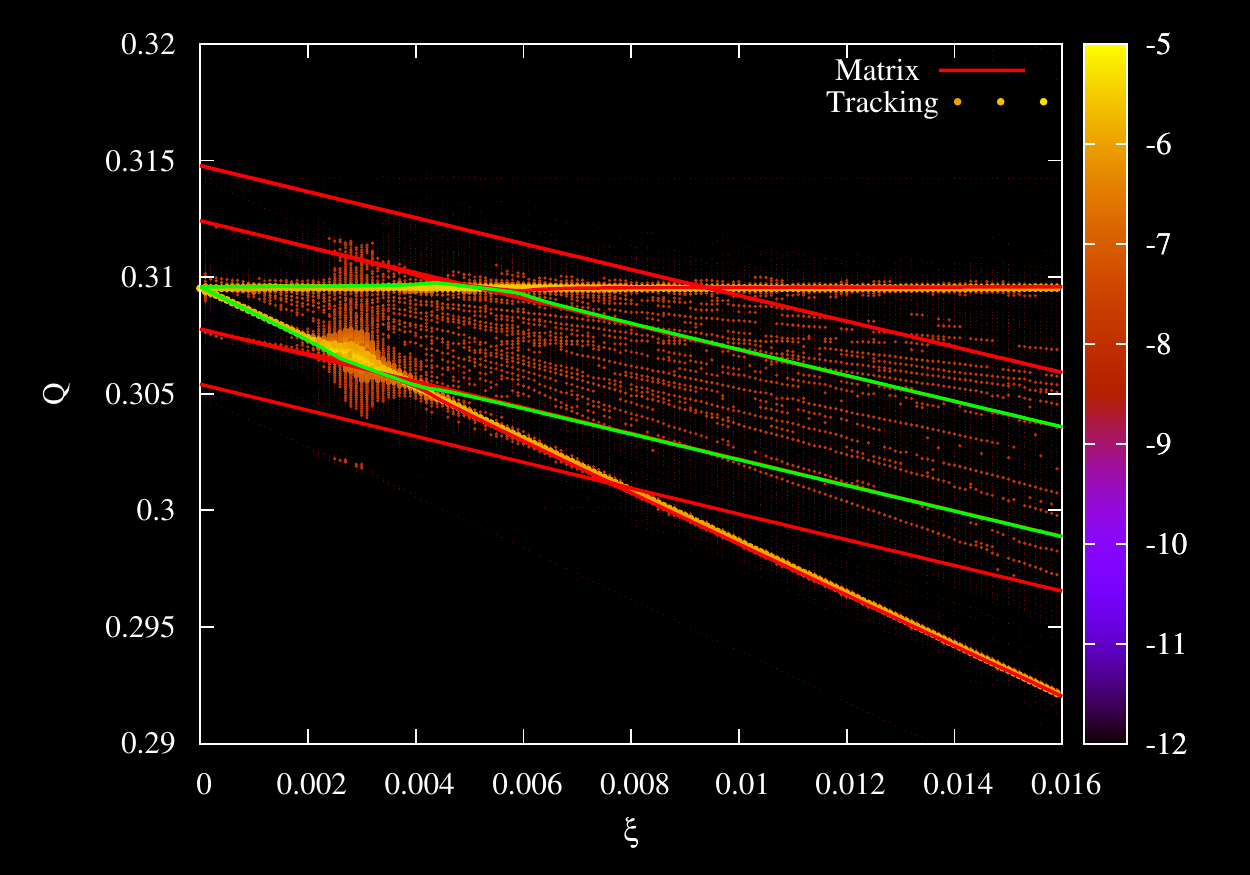}
\end{center}
\caption{Synchro--betatron modes as a function of the beam--beam parameter for $Q'=0.0$ and $\beta^*/\sigma_s\approx90$. The colours correspond to the amplitude of the spectral line.
Impedance was set to be constant over the whole scan.}
\label{bbwk_real}
\end{figure}

\begin{figure}[htb]
\begin{center}
\includegraphics[width=0.45\textwidth]{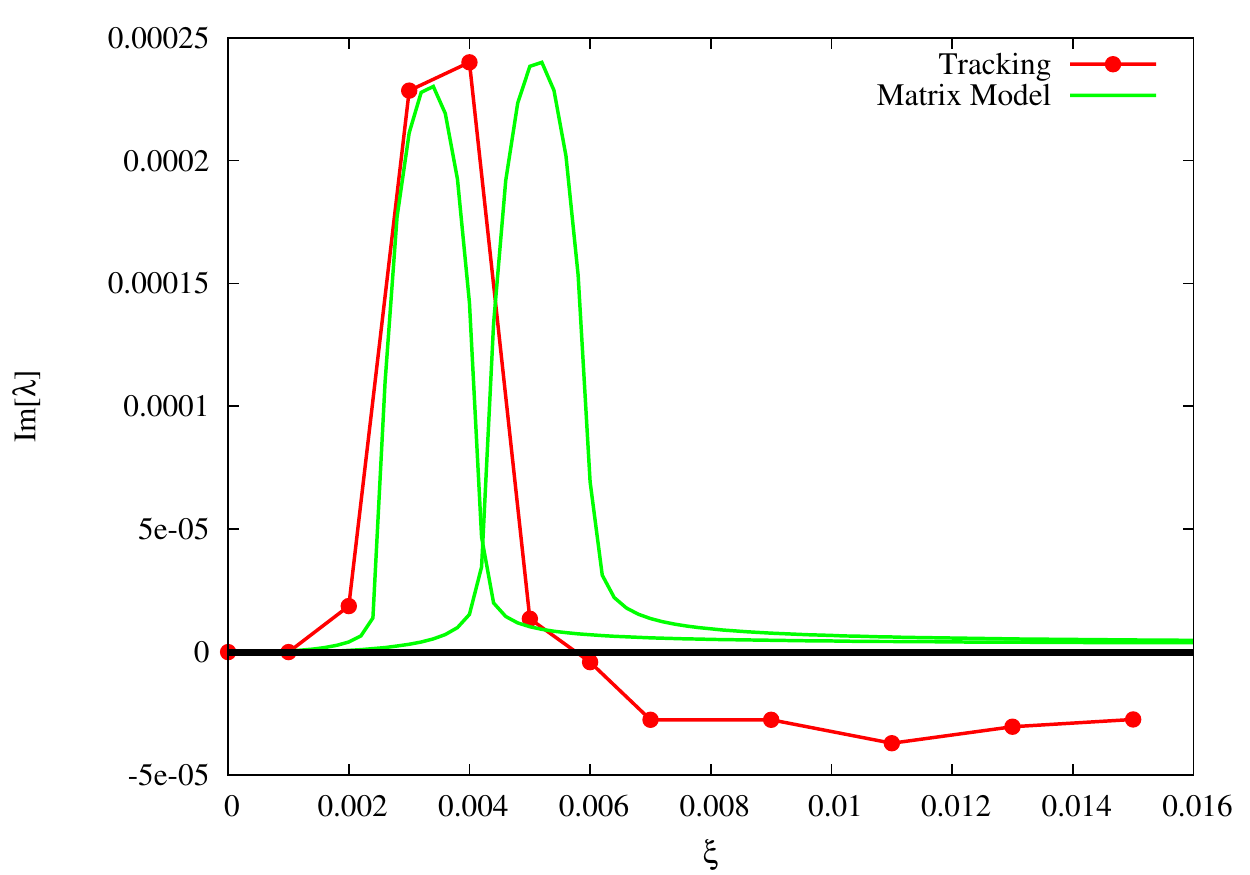}
\end{center}
\caption{Imaginary part of the tune shift of the most unstable modes as a function of the beam--beam parameter for $Q'=0.0$ and $\beta^*/\sigma_s\approx90$.
Impedance was set to be constant over the whole scan.}
\label{bbwk_imag}
\end{figure}

Figures \ref{bbwk_real} and \ref{bbwk_imag} show a scan in the beam--beam parameter at constant impedance. As the beam--beam $\pi$-mode approaches the head--tail mode --1 ($\xi\approx0.003$), they become coupled, leading to a strong instability with similar rise times and characteristics to the impedance-driven TMCI (Transverse Mode Coupling Instability). This is observed in both the tracking and CMM with comparable rise times. The CMM also indicates a coupling between the $\sigma$-mode and the head--tail mode +1. This is not observed in the tracking simulations; the reasons for this discrepancy are under investigation but could be related to Landau damping, which is not taken into account in the CMM. The strength of this coupling instability and the range in terms of $\xi$ over which the modes couple are determined by the strength of the wake and the $\beta$-function at the IP.

\begin{figure}[htb]
\begin{center}
\includegraphics[width=0.45\textwidth]{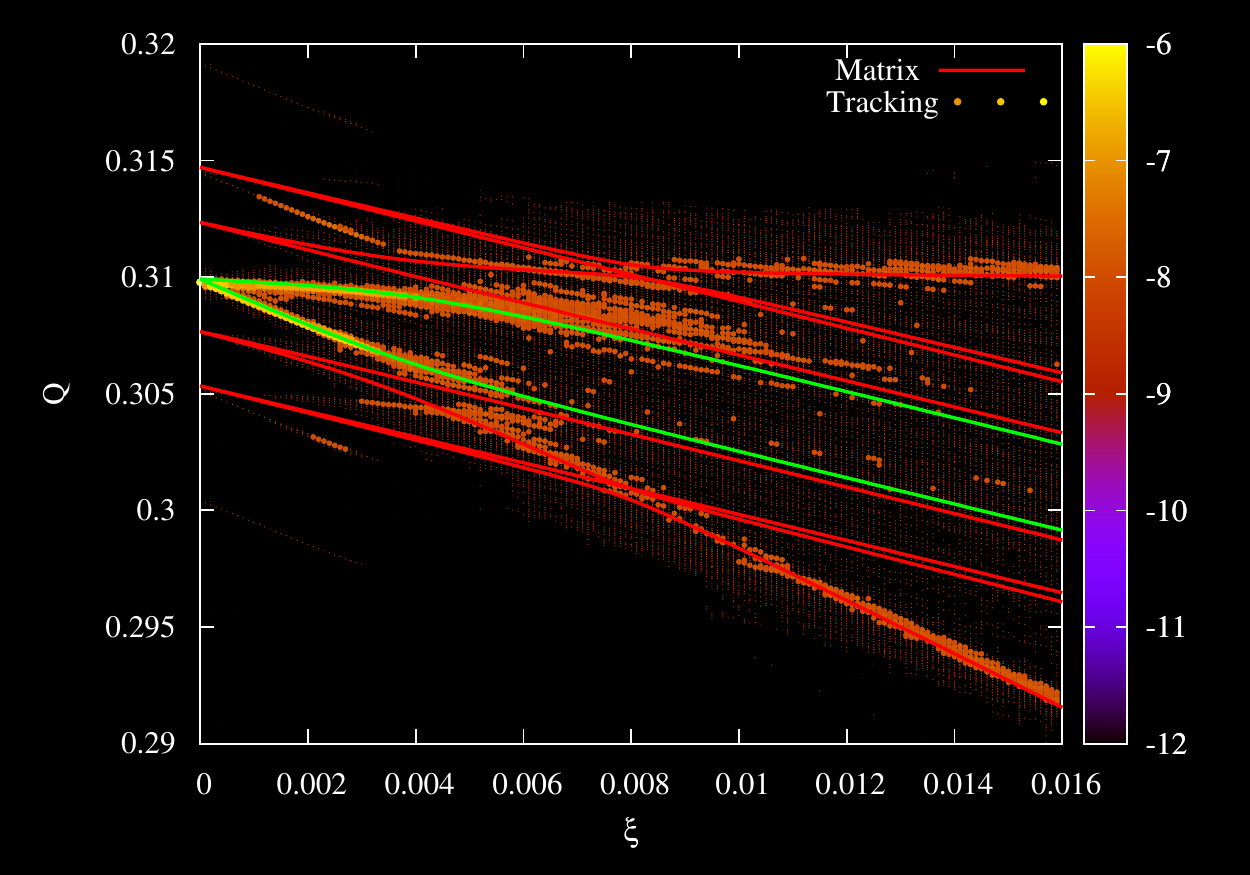}
\end{center}
\caption{Synchro--betatron modes as a function of the beam--beam parameter for $Q'=0.0$ and $\beta^*/\sigma_s\approx1$. The colours correspond to the amplitude of the spectral line.
Impedance was set to be constant over the whole scan.}
\label{bbwk_sbc}
\end{figure}

Figure \ref{bbwk_sbc} shows a scan in the beam--beam parameter using the same beam parameters as in Fig. \ref{bbwk_real} except for the ratio $\beta^*/\sigma_s$. A ratio $\beta^*/\sigma_s\approx1$ introduces synchro--betatron coupling from the beam--beam interaction itself. In this case, as was shown in Fig. \ref{bb6D_1}, the synchrotron sidebands can be deflected by the beam--beam modes. For strong synchro--betatron coupling the most unstable modes involved in the coupling instability (shown in green in Figs. \ref{bbwk_real} and \ref{bbwk_sbc}) are not overlapping any higher order head--tail modes. This results in a suppression of the mode coupling instability observed for higher $\beta^*/\sigma_s$ ratios. Synchro--betatron coupling also increases the Landau damping introduced by the beam--beam interactions. It was shown in Ref. \cite{alexahin} that when the synchrotron tune is of the order of the beam--beam parameter and significant synchro--betatron coupling is present, the tune spread of the lower order sidebands can overlap the $\pi$-mode and damp it. This effect could be reproduced in simulations \cite{sbr} and may be useful in the case where coherent beam--beam mode stability becomes an issue for machines operating at low $\beta^*/\sigma_s$ ratio.

\begin{figure}[htb]
\begin{center}
\includegraphics[width=0.45\textwidth]{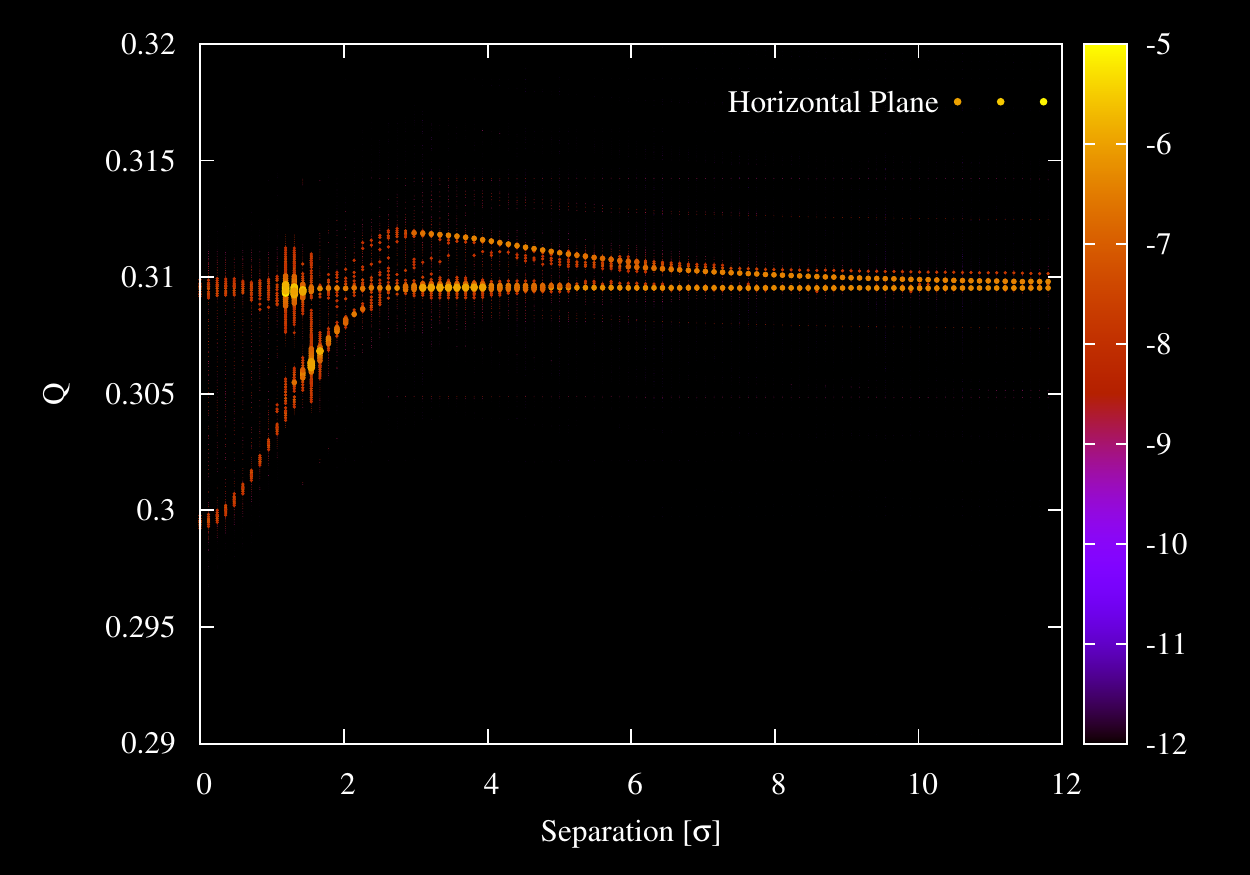}
\end{center}
\caption{Synchro--betatron modes as a function of the transverse separation for $Q'=0.0$ and $\beta^*/\sigma_s\approx90$. The colours correspond to the amplitude of the spectral line.
Impedance and beam--beam parameters were set to be constant over the whole scan.}
\label{bbwk_sep}
\end{figure}

Colliding with transverse offsets changes the frequency of the beam--beam coherent modes and modifies the stability diagram \cite{stability_diag}. Offset collisions can occur while bringing the beam into collisions, in the first moments of a physics store before the luminosity is optimized or when levelling the luminosity, as was routinely done at the LHC in 2012 \cite{leveling}. Figure \ref{bbwk_sep} illustrates a scan in separation (only the separation plane is shown) including coherent beam--beam and impedance. The mode coupling instability is observed when either the $\pi$-mode overlaps the head--tail mode --1 or the $\sigma$-mode overlaps the head--tail mode +1 at separations between 1.0 and 2.0\,$\sigma$. These instabilities also occur when the stability diagram reaches its minimum \cite{stability_diag}. This was tested in a dedicated experiment \cite{lhc_co} during which instabilities were observed at small separations while the beam was fully stable when colliding head-on. Weak--strong simulations with single-plane offset indicate that the stability can be shared between the horizontal and vertical planes preventing any loss of Landau damping and consequently any impedance-only-driven instabilities to rise. This experiment, although not fully conclusive, appears to confirm the existence of the mode coupling instability involving coherent beam--beam modes. Experimental data and detailed analysis can be found in Ref. \cite{lhc_co}.

\begin{figure}[htb]
\begin{center}
\includegraphics[width=0.45\textwidth]{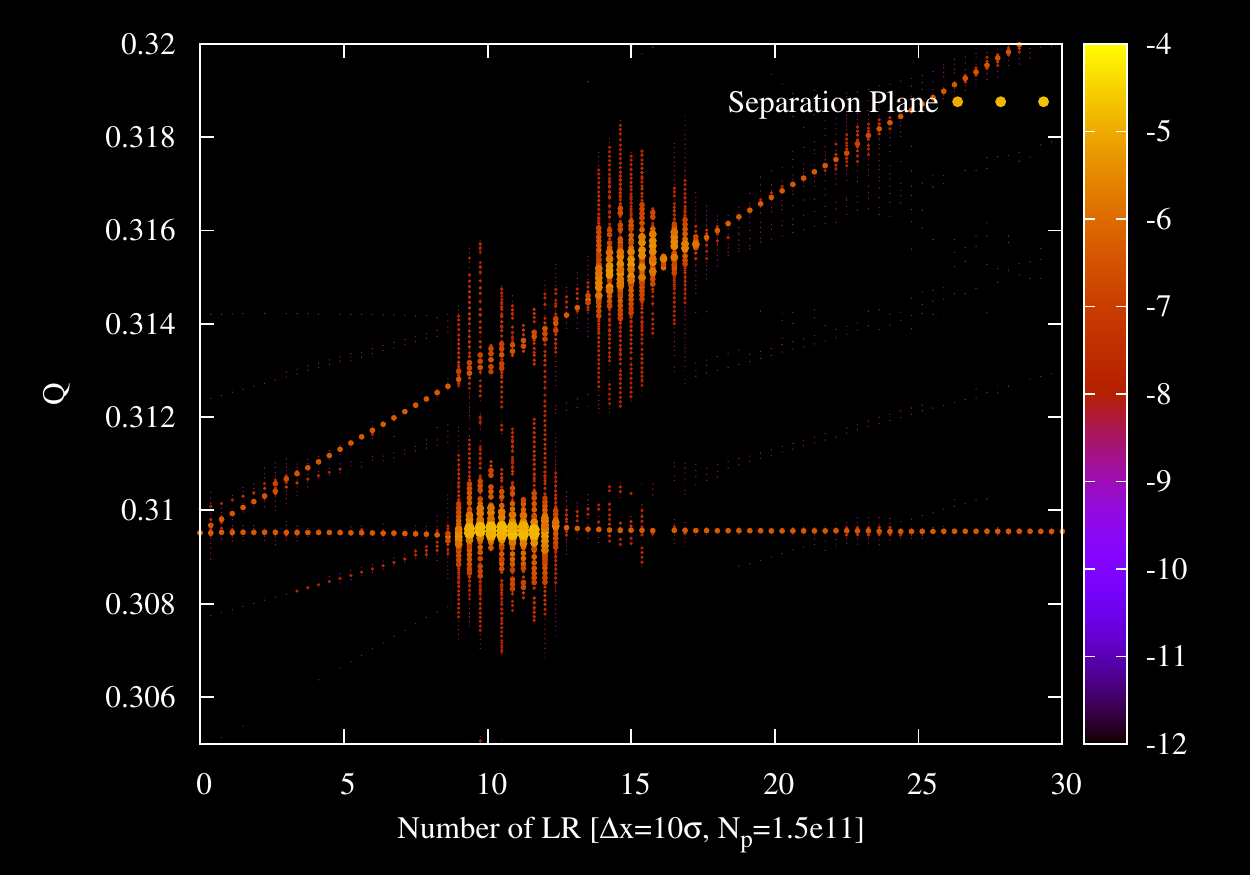}
\end{center}
\caption{Synchro--betatron modes as a function of the number of long-range interactions for a single colliding IP. All the long-range interactions were lumped in one location for which
a separation of 10\,$\sigma$ was assumed.}
\label{bbwk_lr}
\end{figure}

It is worth mentioning that mode coupling instabilities can also occur for long-range interactions when the tune shift is sufficiently high. In the case of the LHC, the $\beta$-functions at the location of the long-range interactions can reach several kilometers, discarding any benefits from synchro--betatron coupling, and this therefore represents the worst-case scenario for this specific mode coupling instability. With the 2012 LHC beam parameters the tune shift at which the instability occurs is reached for approximately 10 long-range interactions, as shown in Fig. \ref{bbwk_lr}. This number has to be compared to 16 long-range interactions per IP in the case of nominal LHC bunches, indicating that PACMAN bunches are the most critical ones. The instability observed for 15 long-range interactions is originating from the other plane.

Figure \ref{bbwk_lr} represents the simplified case of a single IP colliding, where all the interactions were lumped in one place. In reality, multibunch effects and the phase advance between consecutive IPs will modify the situation and should be considered in any realistic simulations of the LHC.

\section{Stabilization of single-bunch instabilities}

Chromaticity combined with tune spread (to provide Landau damping) is generally used to cure transverse instabilities. In the specific case of the LHC, the
bunch-by-bunch transverse damper can also be used for this purpose. In order get a better understanding of how these parameters affect the coherent beam dynamics,
we start with the CMM. All the following simulations were performed using beam parameters corresponding to the most critical configuration, where the transverse mode coupling instability is the strongest.

\begin{figure}[htb]
\begin{center}
\includegraphics[width=0.45\textwidth]{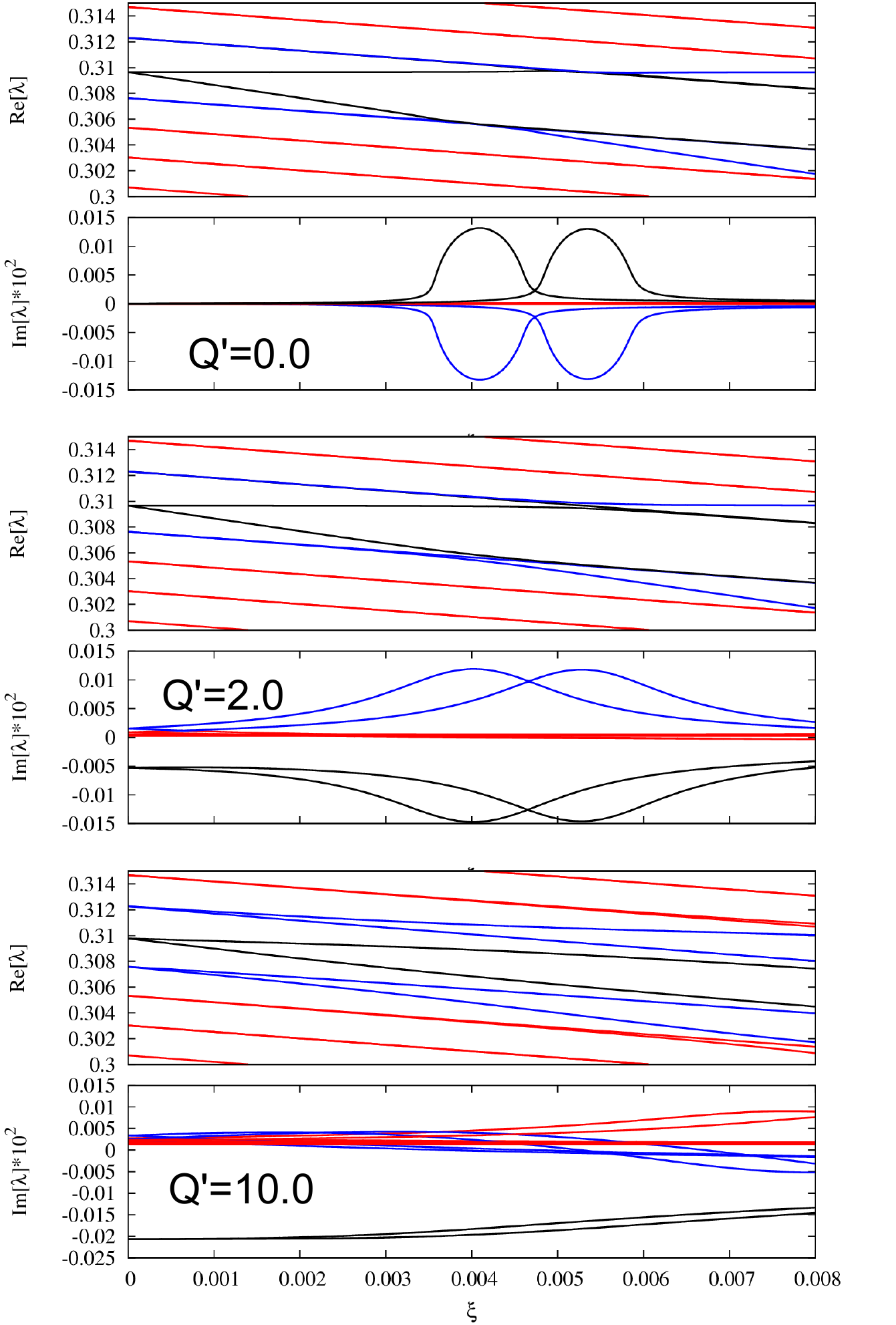}
\end{center}
\caption{Real and imaginary parts of the tune shift as a function of the beam--beam parameters for a single bunch colliding head-on at $Q'=0.0$ (top), $Q'=2.0$ (middle) and $Q'=10.0$ (bottom).
Here the ratio $\beta^*/\sigma_s\approx70$.}
\label{chrom}
\end{figure}

Figure \ref{chrom} shows the real and imaginary tune shifts for a single bunch colliding head-on computed using the CMM for increasing chromaticity. At $Q'=0.0$,
the modes are fully coupled and the instability develops similarly to what was previously shown in Figs. \ref{bbwk_real} and \ref{bbwk_imag}. As the chromaticity is increased
the frequencies of the modes involved in the instability are separated until they fully decouple for $Q'=10.0$. Chromaticity should therefore help mitigating the mode
coupling instability. However, the larger the ratio $\beta^*/\sigma_s$, the higher the chromaticity required to decouple the modes. Operating with too large chromaticity
may degrade beam lifetime. For very large ratio $\beta^*/\sigma_s$, which is typically the case for long-range interactions, using chromaticity only may therefore not
be appropriate to cure this instability.

\begin{figure}[htb]
\begin{center}
\includegraphics[width=0.45\textwidth]{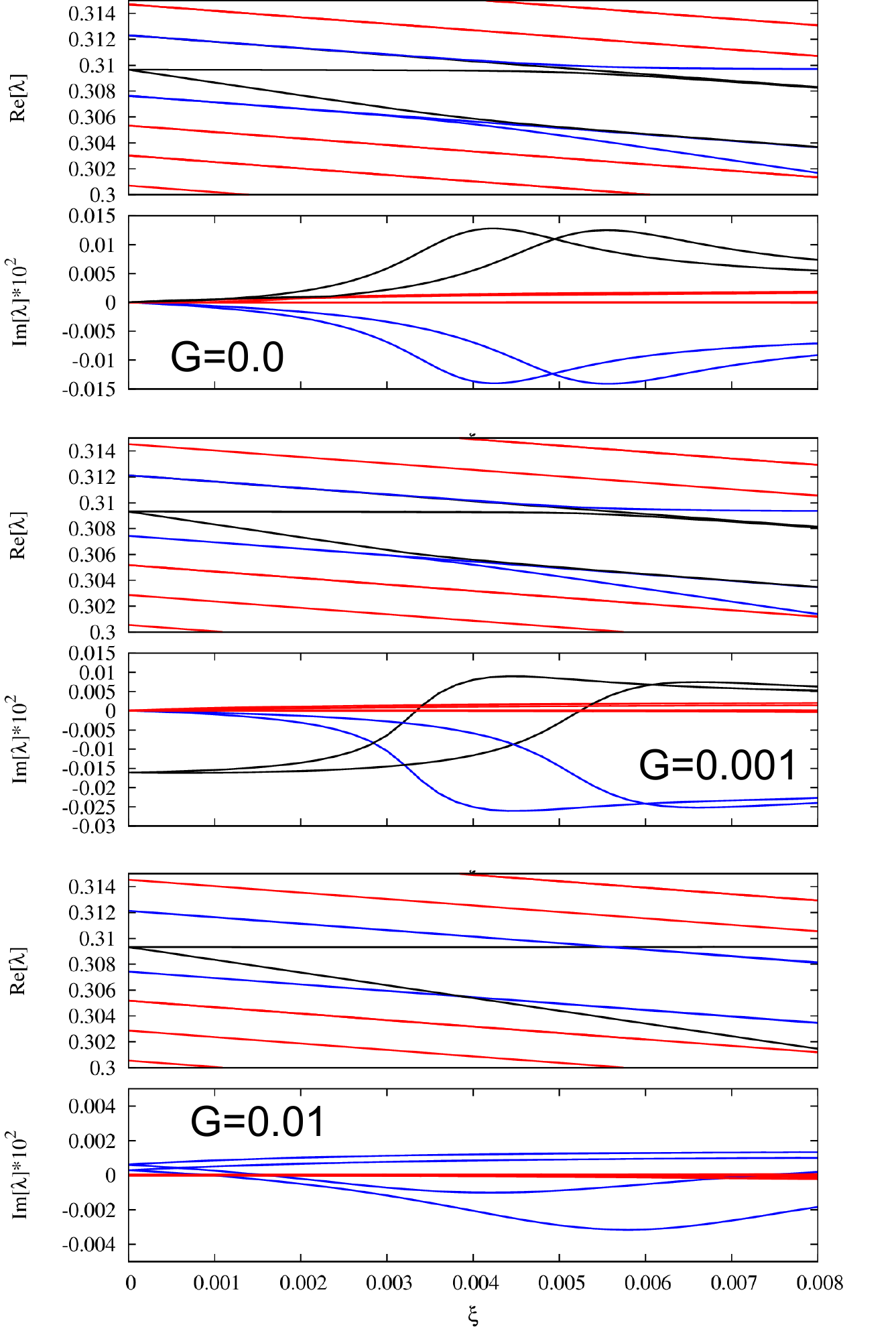}
\end{center}
\caption{Real and imaginary parts of the tune shift as a function of the beam--beam parameters for a single bunch colliding head-on at $G=0.0$ (top), $G=0.001$ (middle) and $G=0.01$ (bottom).
Here the ratio $\beta^*/\sigma_s\approx8$. The gain $G$ is specified in 1/turns.}
\label{damp}
\end{figure}

Figure \ref{damp} shows the real and imaginary tune shifts for a single bunch colliding head-on computed using the CMM for increasing transverse damper gain. The damper is assumed to be an ideal rigid bunch damper for which the gain is defined in 1/turns. The damper is most efficient on modes with a significant dipolar component, such as head--tail mode 0. If Landau damping is sufficient to damp higher order modes, which is generally the case for colliding beams, the transverse damper should be a very efficient means to cure the instability without having to run at unrealistically high gain.

\begin{figure}[htb]
\begin{center}
\includegraphics[width=0.45\textwidth]{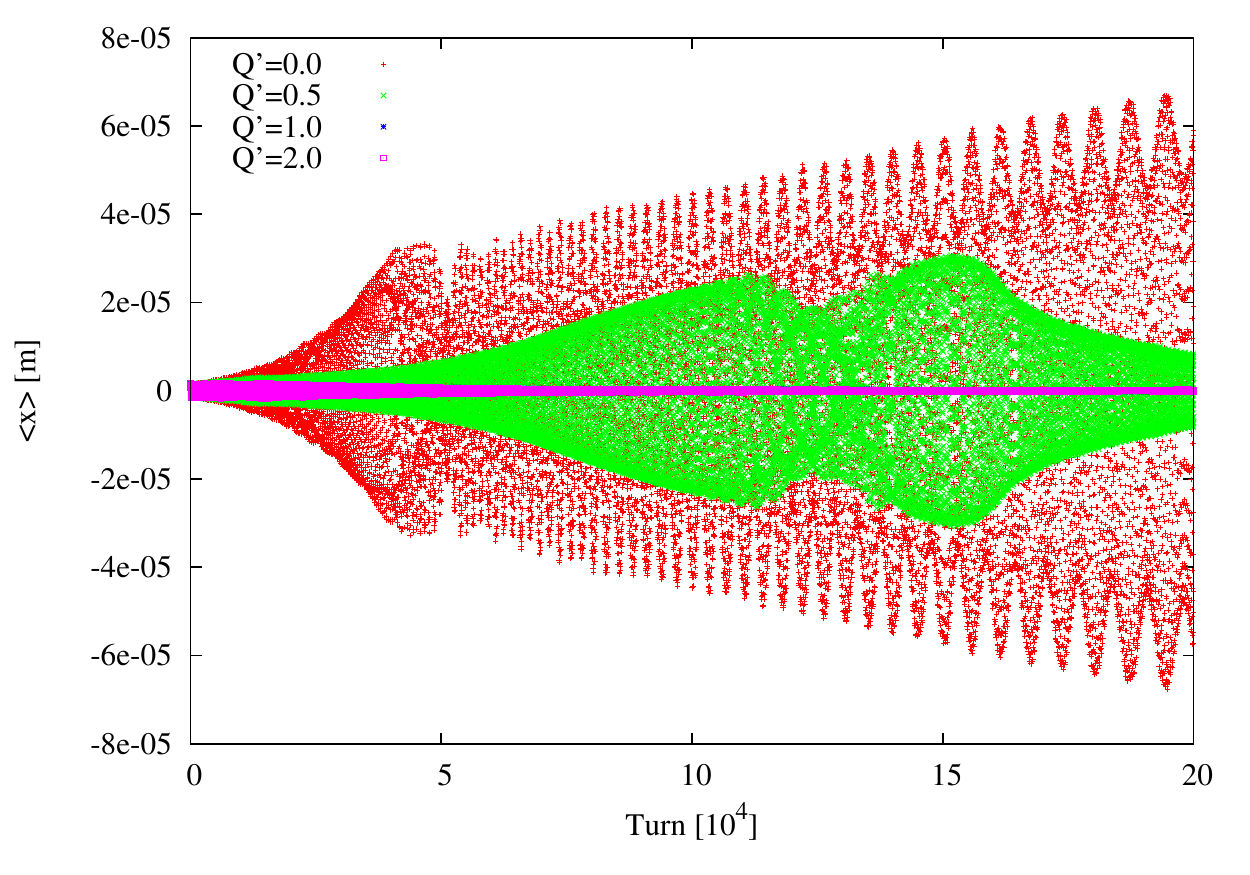}
\end{center}
\caption{Centre of mass motion for a single bunch colliding head-on with $\beta^*/\sigma_s\approx8$ and increasing $Q'$.}
\label{qp_ho_06}
\end{figure}

\begin{figure}[htb]
\begin{center}
\includegraphics[width=0.48\textwidth]{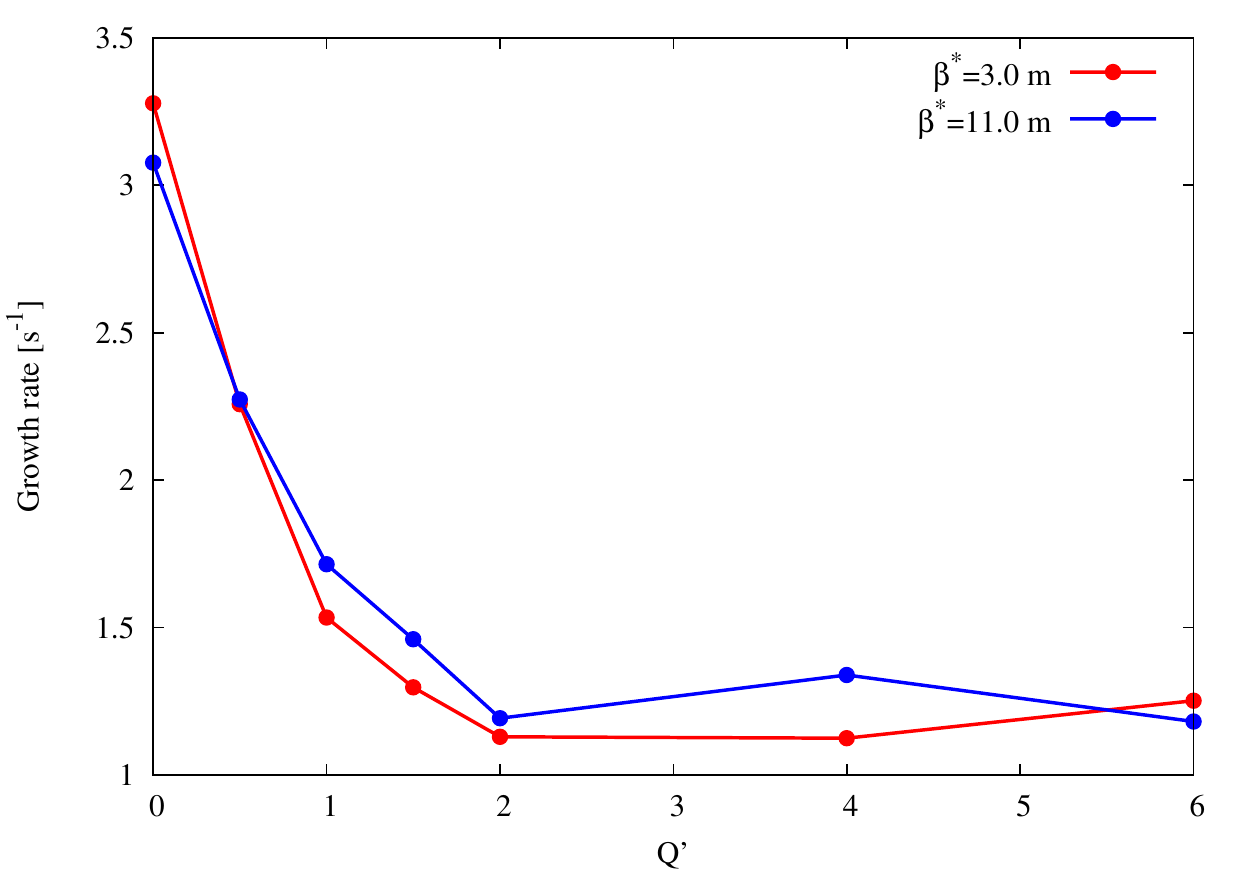}
\end{center}
\caption{Instability rise time for a single bunch colliding head-on with $\beta^*/\sigma_s\approx30$ and $\beta^*/\sigma_s\approx110$ as a function of $Q'$.}
\label{qp_ho_3_11}
\end{figure}

As mentioned before, the CMM does not include Landau damping. In order to assess beam stability including Landau damping, multiparticle simulations are required. Figures \ref{qp_ho_06} and \ref{qp_ho_3_11} show the impact of chromaticity on the mode coupling instability for a single bunch colliding head-on in one IP. For low $\beta^*/\sigma_s$ ratio chromaticity alone is sufficient to damp the instability. As the ratio is increased, chromaticity alone reduces the rise time but does not completely cure the instability up to $Q'=6.0$. Comparing the cases with $\beta^*/\sigma_s\approx30$ and $\beta^*/\sigma_s\approx110$, it seems that the rise time as a function of chromaticity converges for large $\beta^*/\sigma_s$ ratios. This confirms the results from the CMM and indicates possible issues with long-range interactions when using chromaticity only as a cure to the mode coupling instability.

\begin{figure}[htb]
\begin{center}
\includegraphics[width=0.45\textwidth]{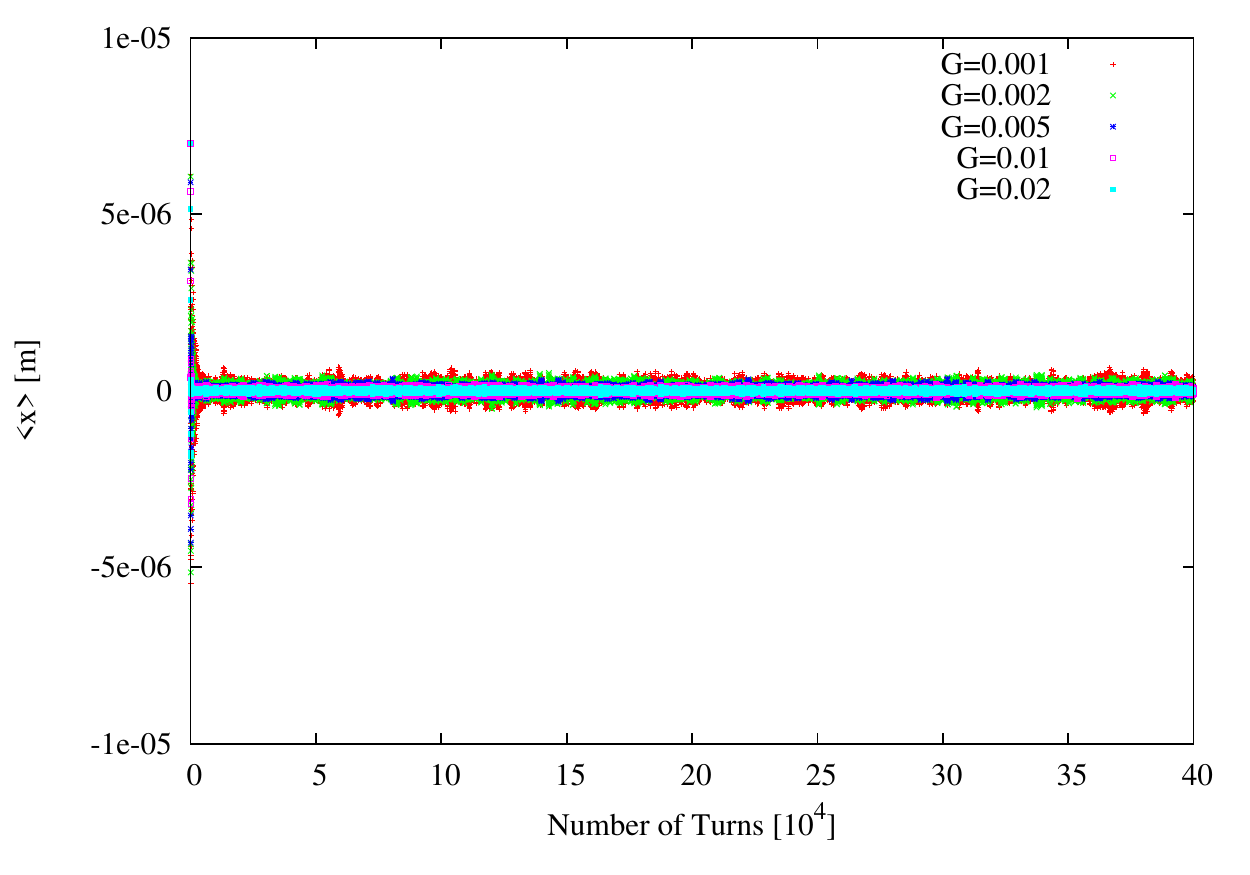}
\end{center}
\caption{Centre of mass motion for a single bunch colliding head-on with $\beta^*/\sigma_s\approx100$ and $Q'=0.0$ as a function of the damper gain.}
\label{g_ho}
\end{figure}

Figure \ref{g_ho} shows the impact of the transverse damper on the mode coupling instability for a single bunch colliding head-on in one IP with $\beta^*/\sigma_s\approx100$ and $Q'=0.0$. It is seen that even for a very low gain of 0.001 the beam is rendered stable by the transverse damper. This benefit of the transverse damper was experimentally demonstrated with offset collisions. It was shown that the beams were strongly unstable when the damper was turned off and stable with the damper on for a separation of approximately 1.0\,$\sigma$ \cite{lhc_co} and with $Q'\approx5.0$. This strong instability with offset collisions and damper off is a good candidate for the mode coupling instability mentioned before.

\begin{figure}[htb]
\begin{center}
\includegraphics[width=0.45\textwidth]{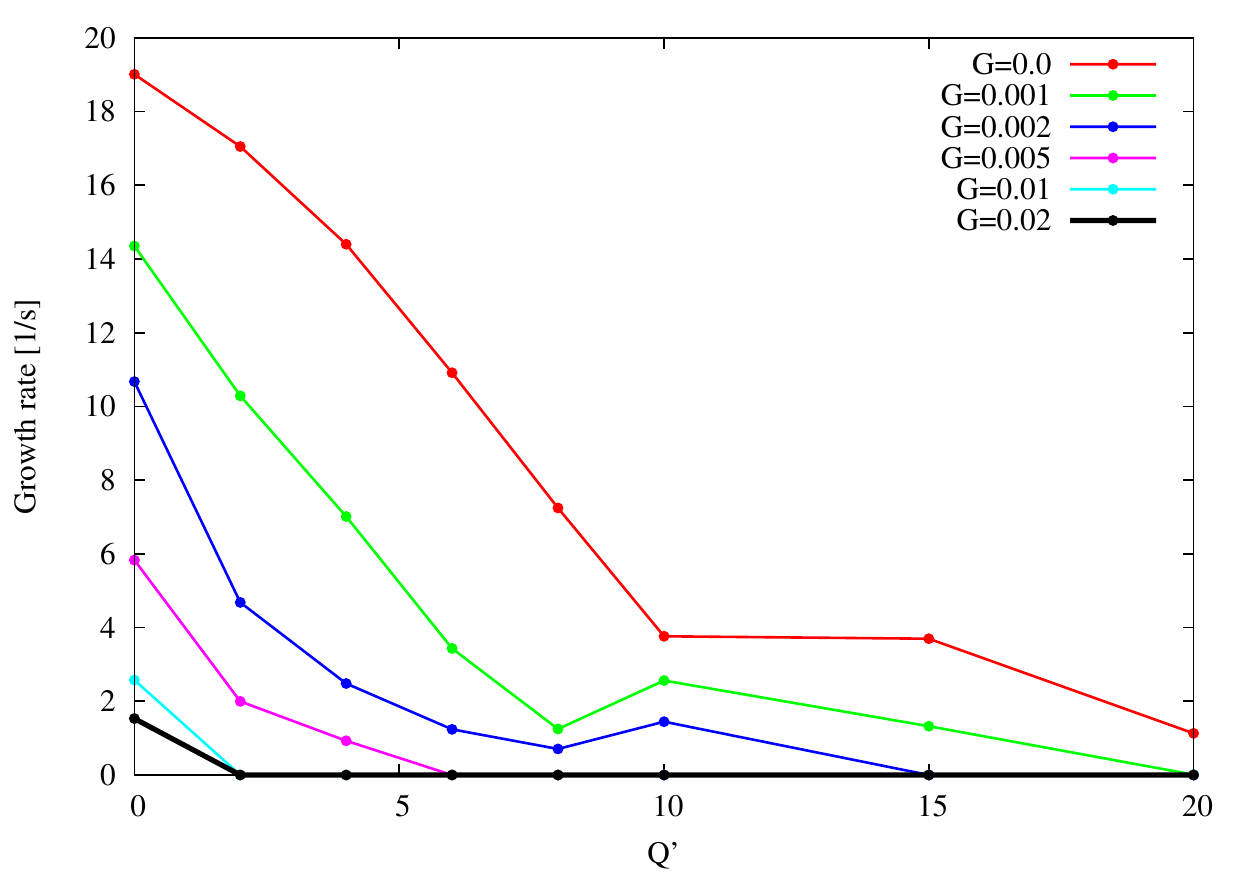}
\end{center}
\caption{Instability rise time as a function of chromaticity for different damper gains, in the single-bunch approximation of a configuration with 10 long-range interactions separated by 10\,$\sigma$.}
\label{g_qp_lr}
\end{figure}

Although it seems that any instability involving head-on interactions could be cured with the transverse damper, the situation is more complicated when looking at offset collisions and even more for long-range interactions. Two mechanisms can degrade the situation: the reduction of the tune spread and the absence of synchro--betatron coupling in the case of long-range interactions. Figure \ref{g_qp_lr} shows the instability rise time as a function of transverse damper gain and chromaticity in the case of long-range interactions. All the interactions are lumped in one location and the tune shift was set to be equivalent to 10 long-range interactions with a separation of 10\,$\sigma$. In this case, either chromaticity or damper only is not sufficient to damp the instability. There is however a correlated dependence: the higher the gain the lower the chromaticity required to stabilize the beams and inversely. In 2012, the LHC was operated with both high gain and chromaticity. Both these parameters are known to degrade beam lifetime and emittance; these results indicate that there should be room for optimization of these parameters which should be considered during LHC recommissioning in 2015.

Adding a full head-on interaction significantly increases the tune spread and consequently damps all long-range types of instabilities. This is consistent with experimental observations \cite{obs_LHC} and indicates that operating the transverse damper during physics stores may not be required as long as sufficient tune spread is provided and tune shifts at which the coupled mode instability occurs are avoided.

\section{Multibunch effects}

Previous results were produced using the single-bunch approximation. In the LHC, the filling pattern, long-range interactions and symmetry of the collision points should be considered to give a more realistic picture of the coherent beam dynamics. In such configuration, each bunch encounters a different set of beam--beam interactions, leading to significantly different incoherent and coherent effects. In particular, the coupling between bunches along a single beam through the machine impedance and to bunches of the other beam through beam--beam interactions becomes different for each bunch. The CMM was extended to multibunch in order to properly take into account the real beam--beam configuration and coupled bunch impedance.

\begin{figure}[htb]
\begin{center}
\includegraphics[width=0.45\textwidth]{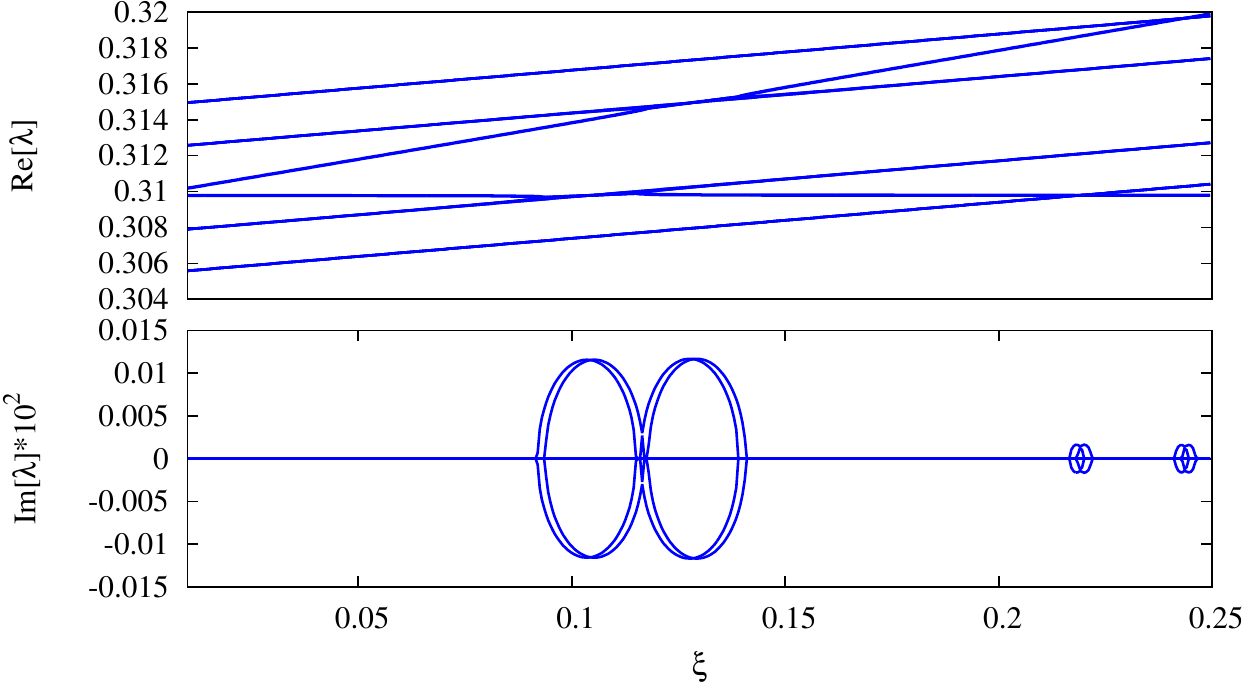}
\includegraphics[width=0.45\textwidth]{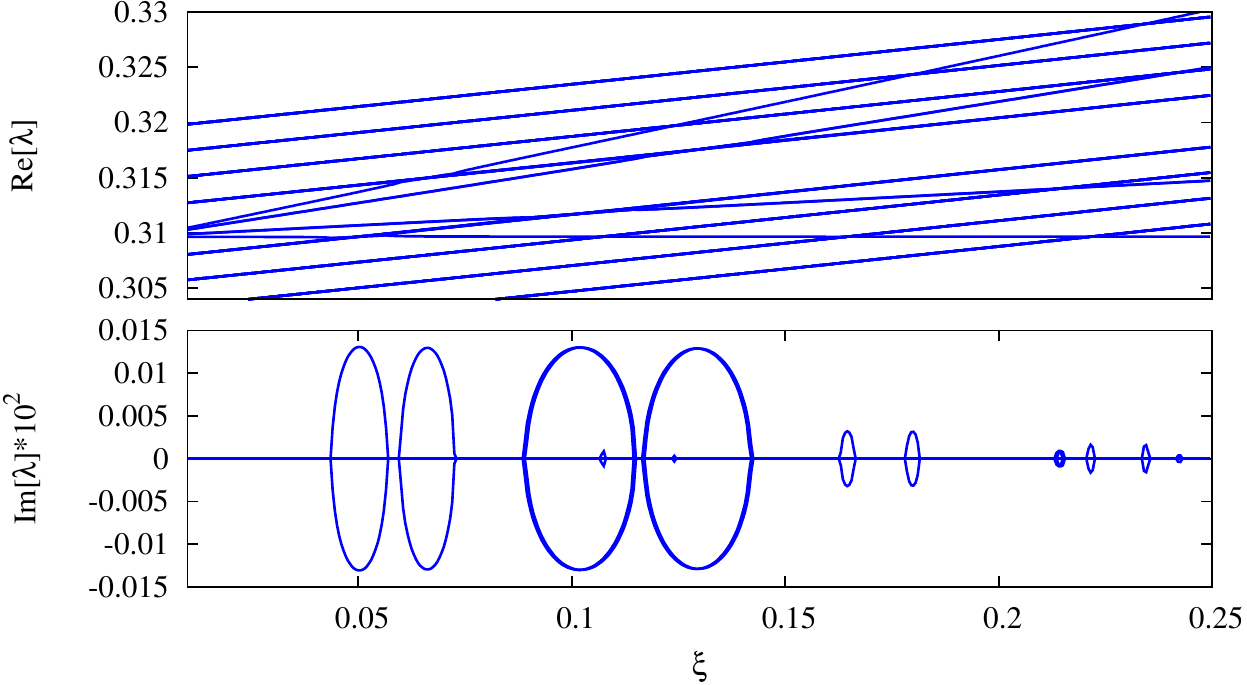}
\includegraphics[width=0.45\textwidth]{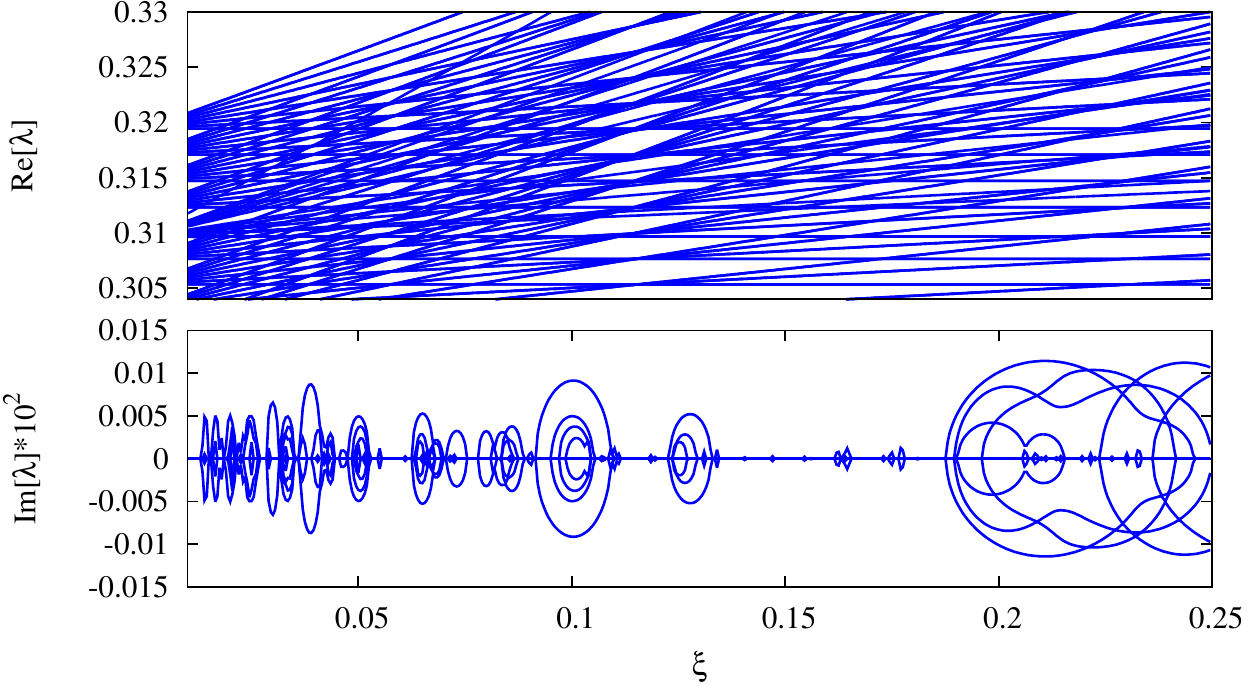}
\end{center}
\caption{Real and imaginary parts of the tune shift as a function of the beam--beam parameter. In this case only long-range interactions in a single IP were considered.
The top plot is for two bunches in each beam, the middle plot for three bunches in each beam and the bottom plot for eight bunches in each beam colliding only on one side of the IP
to enhance PACMAN effects. The modes of all bunches are shown in these plots.}
\label{mb_circ}
\end{figure}

\begin{figure}[htb]
 \centering
\includegraphics[width=1.0\linewidth]{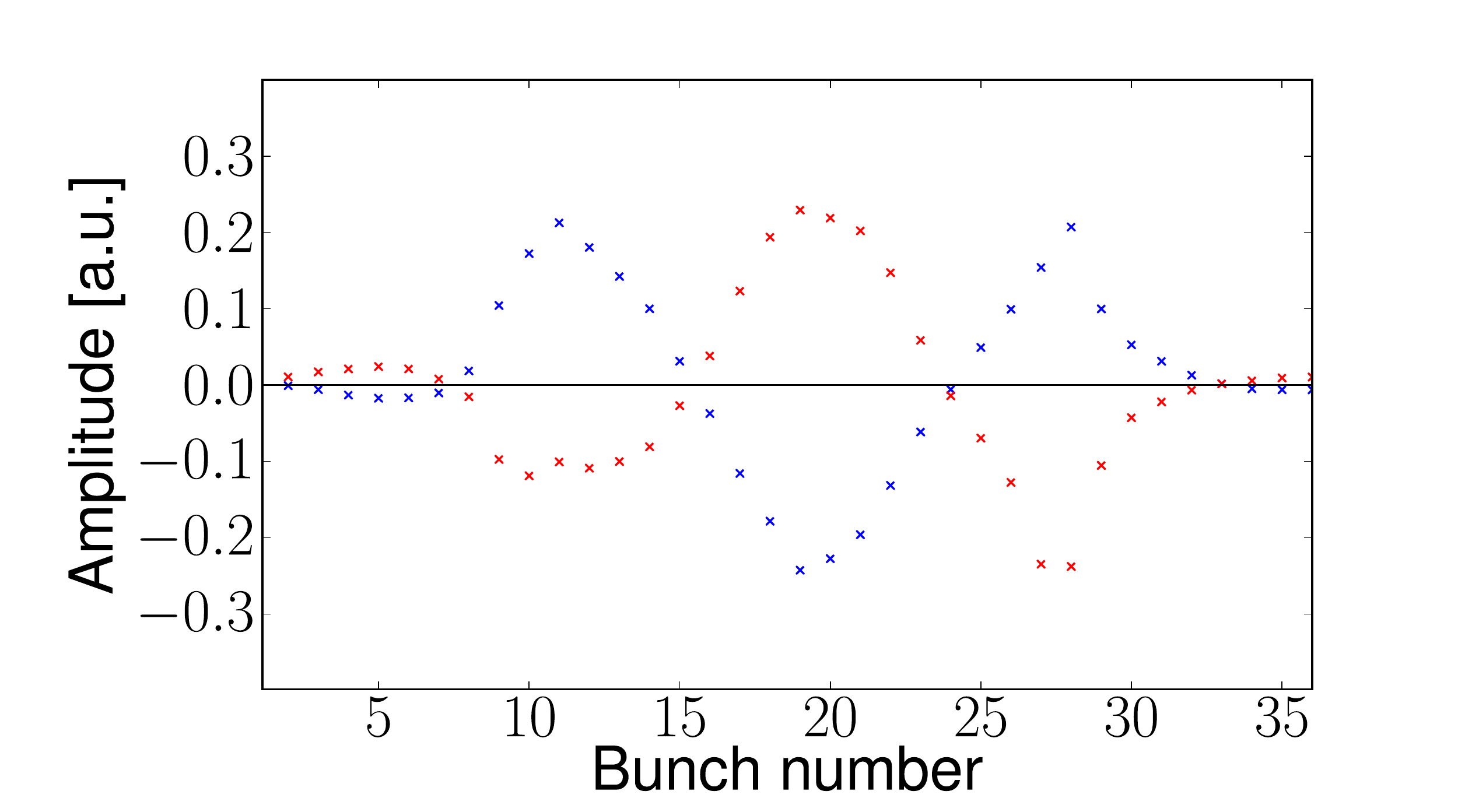}
\includegraphics[width=1.0\linewidth]{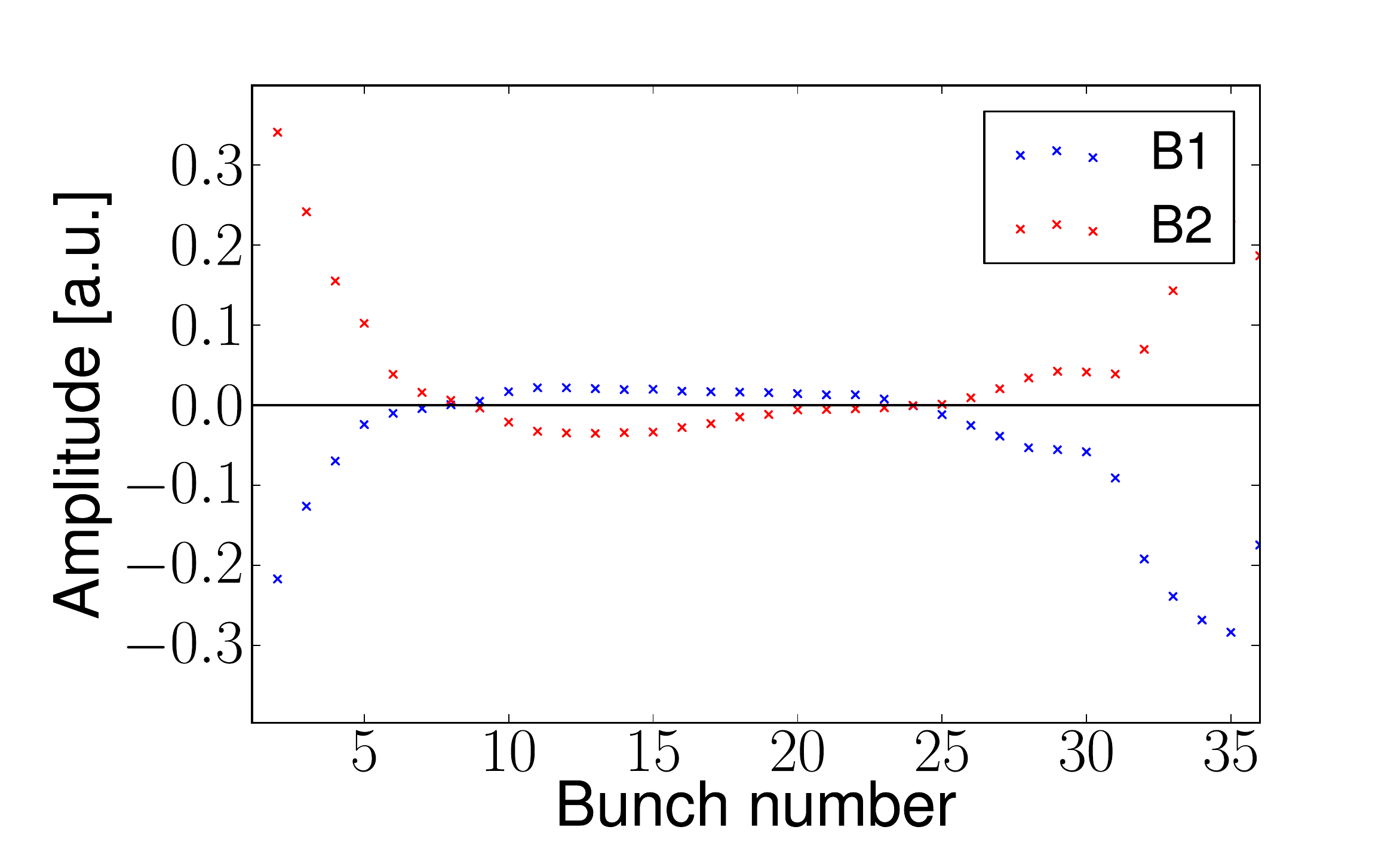}
\caption{Oscillation amplitude of each bunch of the two beams for an unstable mode in the presence of impedance and 16 long-range beam--beam interactions in one IP with separations of 10\,$\sigma$ (upper plot) and 9\,$\sigma$ (lower plot).}
\label{multibunch_eigenmode}
\end{figure}

Figure \ref{mb_circ} shows how the complexity of the collision pattern can modify the coherent beam dynamics. In these cases only long-range interactions in a single IP were considered. The top plot shows the case of two bunches per beam, each bunch colliding only once on either side of the IP, and is therefore similar to the single-bunch case. The only difference between the bunches comes from the coupled bunch impedance. The coupled mode instabilities are clearly observed when the beam--beam modes cross the head--tail modes $\pm$1. Although much weaker, this instability is also observed when the beam--beam modes cross the head--tail modes $\pm$2. This is an interesting feature which deserves further investigation, as the effect of the damper on such higher order mode coupling instabilities is unclear. As the number of bunches is increased, and hence the complexity of the collision pattern, a lot of modes with different frequencies appear and consequently mode coupling instabilities occur for most beam--beam parameters. In particular, the different modes do not necessarily involve the whole beam, but only a subset of bunches, as also shown by self-consistent simulations in Fig. \ref{multibunch_eigenmode}. It is rather clear from this picture that PACMAN effects and the overall complexity of the LHC collision pattern cannot be neglected when looking at coherent beam dynamics. \\

\begin{figure}[htb]
\begin{center}
\includegraphics[width=1.0\linewidth]{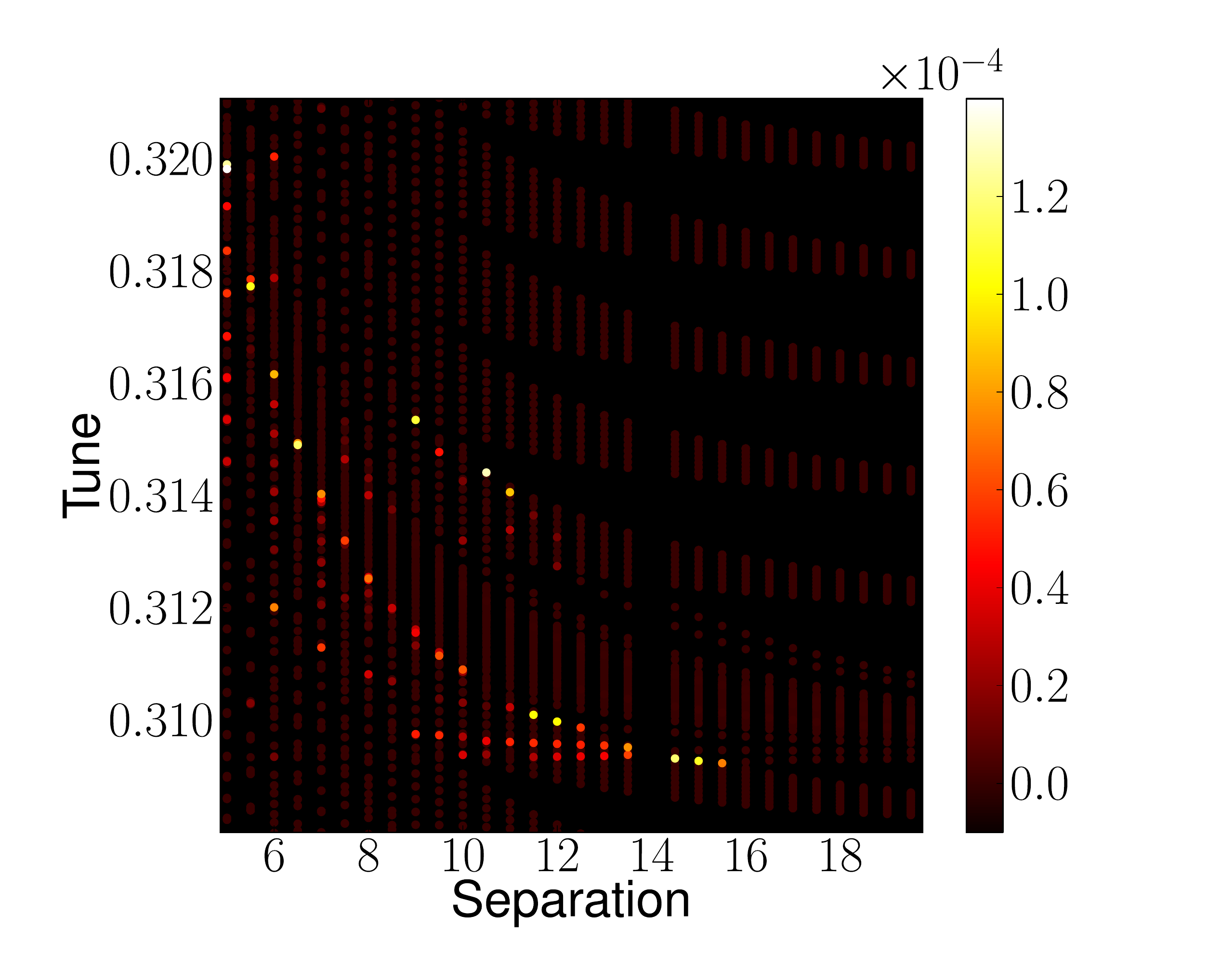}
\includegraphics[width=1.0\linewidth]{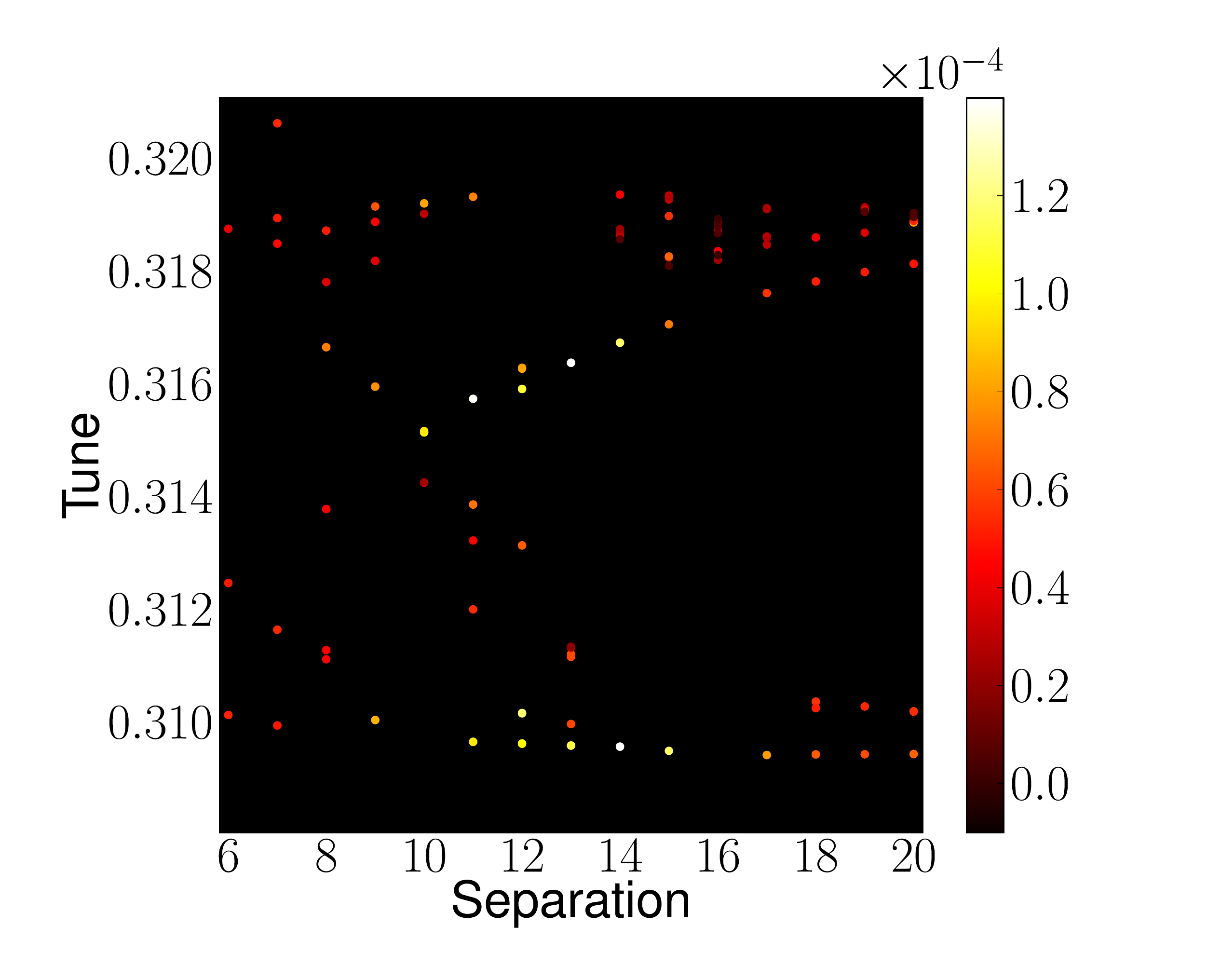}
\end{center}
\caption{Coherent mode frequencies and rise times (colour) as a function of the normalized separation in the horizontal plane at the long-range location for two trains of 36 bunches colliding in one IP with 16 long-range interactions. The upper plot shows the result from CMM and the lower from COMBI. The synchrotron tune in this case was set to 0.002.}
\label{sep_scan}
\end{figure}

Figure \ref{sep_scan} shows a comparison of simulations with the CMM and COMBI in an identical configuration (nominal 2012 LHC running condition with $Q_s=0.002$), indicating a good agreement between the models. In particular, mode coupling instabilities are observed for separations from 11 to 16\,$\sigma$ at frequencies around the horizontal tune, 0.31. The separation
at which the instabilities occur depends on the beam--beam parameter and synchrotron tune. The instabilities observed in the tracking code only, at frequencies between 0.316 and 0.325, originate from the vertical plane and therefore do not appear in the CMM, which models one plane only. It is foreseen to further extend the CMM to take this into account. \\
The complex tune shift of each mode is evaluated by singular value decomposition of the turn-by-turn position of each bunch. The singular vectors associated with the most unstable mode at separations of 9 and 10\,$\sigma$ are shown in Fig. \ref{multibunch_eigenmode}. In particular, one observes that, in the configuration with 10\,$\sigma$ separation, bunches at the edge of the train are stable, while the bunches at the centre of the train are unstable. The opposite is true at 9\,$\sigma$. Moreover, the rise time associated with these two modes are significantly different, respectively 1800 and 3800 turns. This observation provides another indication of the importance of PACMAN effects on the coherent dynamics and therefore motivates further studies of the effect of the different stabilization techniques in configurations as close as possible to the real LHC.

\section{Summary and outlook}

During the 2012 proton run of the LHC, coherent instabilities were routinely observed in the LHC at the end of the squeeze and with beams colliding with a transverse offset, triggering a renewed interest to pursue studies of the interplay of beam--beam and impedance. For this purpose two models were developed:

\begin{itemize}
 \item An analytical model based on the circulant matrix approach used in Ref. \cite{ht_bb_circ}
 \item A fully self-consistent multiparticle tracking model. Single-bunch effects were studied with the code BeamBeam3D \cite{BB3D} and multibunch effects
       with the code COMBI \cite{pieloni_PhD}
\end{itemize}

A full benchmarking campaign with existing code was done to validate the implementation of impedance in these different models and excellent agreement was reached for both the analytical model and tracking codes.

Single-bunch effects were studied in detail for various cases. It was shown that the coherent beam--beam modes can couple with higher order head--tail modes giving rise to strong instabilities with similar characteristics to the impedance-driven TMCI. Possible cures were considered and it was demonstrated that, in the case of the single-bunch approximation, a combination of chromaticity and transverse damper should stabilize the beams in all cases. Specific cases should be studied in detail to optimize the values of the gain and chromaticity. Nevertheless, due to its complex collision pattern, the LHC beams can hardly be approximated with a single bunch. PACMAN effects and coupled bunch impedance have to be considered in any attempt to realistically model the LHC. Both analytical and tracking models were developed to study multibunch effects. Preliminary results show good agreement and tend to confirm the invalidity of the single-bunch approximation for the case of the LHC. Further efforts should be pursued in this direction to provide a better understanding of LHC observations.

\section{Acknowledgments}

The authors would like to thank T. Pieloni and N. Mounet for their help and support regarding these studies.

\end{document}